\documentclass[aps,pra,onecolumn,11pt,showpacs,superscriptaddress,amsmath,amssymb,nobibnotes]{revtex4-1}

\usepackage{graphicx}
\usepackage{amsfonts}    
\usepackage{graphicx}   
\usepackage{verbatim}   
\usepackage{color}      
\usepackage{subfigure}  
\usepackage{hyperref}   
\usepackage{natbib}
\usepackage{booktabs}
\usepackage{threeparttable}
\usepackage{dcolumn}
\usepackage{multirow}

\begin{document}
\title{Quantum-Inspired Gravitational-Wave Detector}
\author{Meng-Jun Hu}\email{mengjun@mail.ustc.edu.cn }
\author{Yong-Sheng Zhang}\email{yshzhang@ustc.edu.cn (corresponding author)}
\affiliation{Laboratory of Quantum Information, University of Science and Technology of China, Hefei 230026, China}
\affiliation{Synergetic Innovation Center of Quantum Information and Quantum Physics, University of Science and Technology of China, Hefei 230026, China}
\date{\today}

           Essay written for the Gravity Research Foundation 2018 Awards for Essays on Gravitation

\begin{abstract}
A new gravitational-wave detector, which is devised based on quantum weak measurement amplification, is introduced and shown has the potential to significantly improve the strain sensitivity of gravitational-wave detection.

\end{abstract}

\maketitle
Gravitational waves have been detected by Laser Interferometer Gravitational-Wave Observatory (LIGO) in 2016, opening the era of gravitational-wave astronomy \cite{gw1,gw2,neutron}. The further dramatic improvements of strain sensitivity of LIGO will require significant technology development and new facilities \cite{gw3}. However, the recent rapid development of quantum weak measurement theory, on the other hand, provides us with the opportunity of designing a totally new gravitational-wave detector \cite{hu}. The quantum-inspired gravitational-wave detector, which operates based on quantum weak measurements amplification, will be introduced here and shown has the potential to significantly improve the strain sensitivity with only current technologies and facilities. 

Quantum weak measurement theory was proposed by Aharonov, Albert, and Vaidman in 1988 \cite{AAV} and has attracted extensive interests in the last decade for its power in solving quantum paradox, reconstructing quantum states and, especially, amplifying ultra-small signal \cite{wv}. The quantum weak measurement theory focus on the situation in which the coupling between the system and the pointer is extremely weak such that almost no information of the system can be obtained. The operation of $\it{post-selection}$ is applied after the weak coupling to pick out system with the definite initial state and the final state. Surprisingly, the signal (e.g., phase) to be measured can be significantly amplified if the post-selection is properly made. The amplified signal can be easily extracted by performing suitable measurement on the pointer after post-selection. The post-selection, which is the key of quantum weak measurement, acts like a sifter to sift out the amplified signal, although at the price of low successful probability. 

Fig. 1 shows the schematic diagram of the gravitational-wave detector designed based on quantum weak measurement amplification. The quantum-inspired gravitational-wave detector consists of five parts i.e., laser source, initial state preparation, gravitational-wave signal collection, signal amplification and signal detection.

The laser source produces high stabilized light beam with polarization $|+\rangle=(|H\rangle+|V\rangle)/\sqrt{2}$, where $|H\rangle$ and $|V\rangle$ represent horizontal and vertical polarization respectively. After passing through a $50:50$ beam splitter (BS1), the state of photons becomes $(|u\rangle+|d\rangle)/\sqrt{2}\otimes|+\rangle$ with $|u\rangle$ and $|d\rangle$ denote path state of photons along up and down arm respectively. The gravitational-wave signal collection is completed by two symmetrical placed polarization Michelson interferometers (PMI) with arm cavities and dual power recycling. Each PMI is a mirror imagine of the other such that differential change of two PMIs can be detected. In practical design, we can fold the setup such that two PMIs share the same vacuum pipelines and test masses. 
Suppose that a gravitational wave with polarization $h_{+}$ is passing through the gravitational-wave detector that placed perpendicular to its direction of propagation, then one arm of PMI is stretched whilst the other one is contracted and the relative length change of each arm is $\bigtriangleup L=hL/2$. The state of photons, after passing through two PMIs, reads
\begin{equation}
|\Psi\rangle=\dfrac{1}{\sqrt{2}}[|u\rangle\otimes\dfrac{1}{\sqrt{2}}(e^{i\theta}|H\rangle+e^{-i\theta}|V\rangle)+|d\rangle\otimes\dfrac{1}{\sqrt{2}}(e^{-i\theta}|H\rangle+e^{i\theta}|V\rangle)],
\end{equation}
where $\theta=2G_{p}G_{arm}k\bigtriangleup L$ with $G_{p}$ and $G_{arm}$ are gains of power recycle and Fabry-Perot arm respectively. Note that global phase $G_{p}G_{arm}kL$ is omitted here. 

\begin{figure}[tbp]
\centering
\includegraphics[scale=0.62]{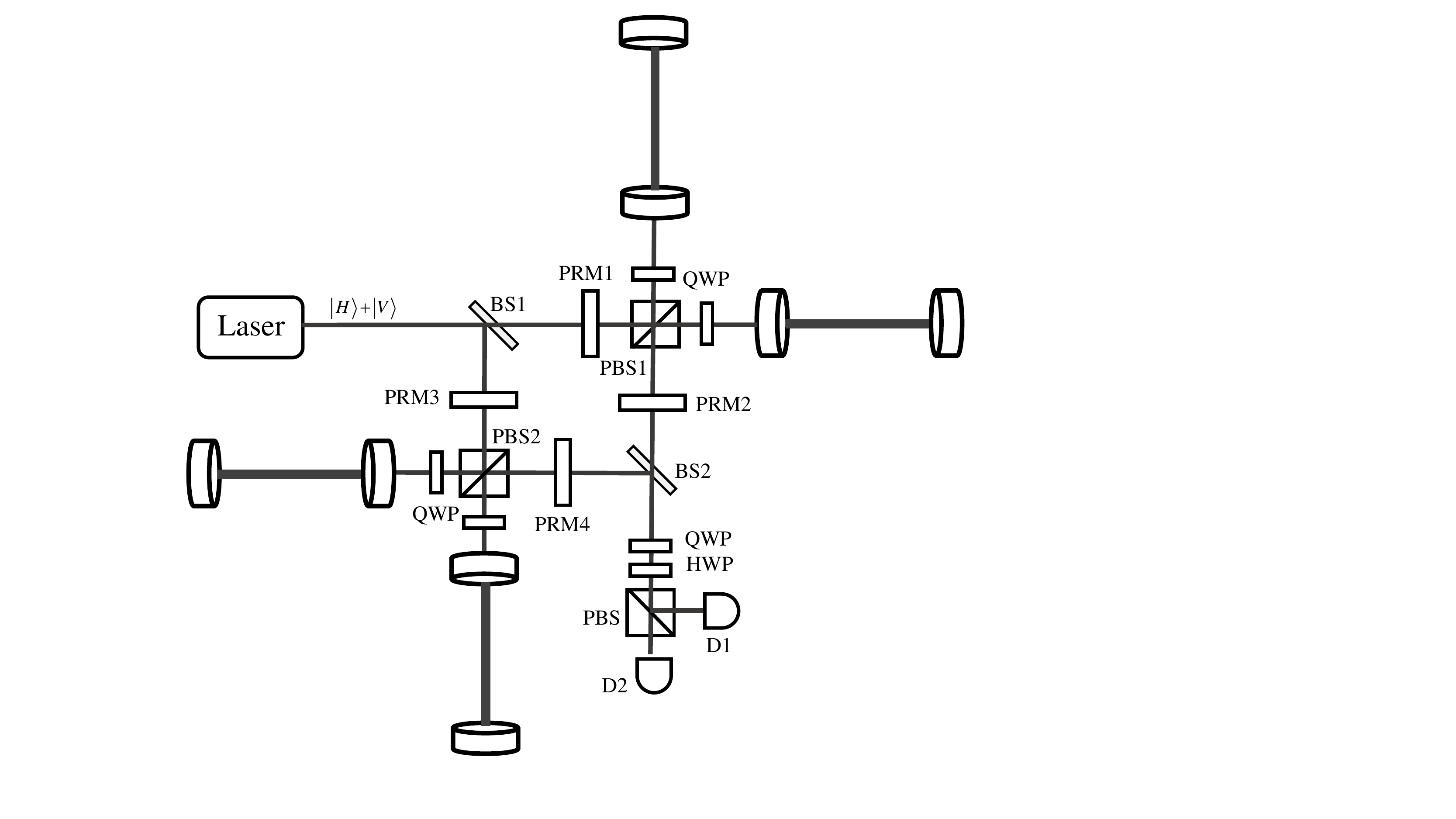}
\caption{Schematic diagram of Quantum Weak Measurements Amplification based Laser Interferometer Gravitational-wave Detector. }
\end{figure}

Signal amplification is fulfilled by post-selection of path state of photons, which is realized by only collecting photons from the dark port of the BS2 as shown in Fig. 1. The post-selected path state of photons is $|\varphi\rangle=t|u\rangle-r|d\rangle$ with $t$ and $r$ are real coefficients of transmission and reflection respectively. The pointer i.e., polarization state of photons, after post-selection, becomes (unnormalized)
\begin{equation}
|\tilde{\phi}\rangle=\langle\varphi|\Psi\rangle=e^{i\theta}(t-re^{-i\cdot 2\theta})|H\rangle+e^{-i\theta}(t-re^{i\cdot 2\theta})|V\rangle.
\end{equation}
Since $\theta\ll 1$ in the case of gravitational-wave detection, $t-re^{i\cdot 2\theta}=(t-r)e^{i\gamma}$ in the first order approximation, the normalized state of polarization then reads
\begin{equation}
|\phi\rangle=\dfrac{1}{\sqrt{2}}(|H\rangle+e^{i\cdot 2(\gamma-\theta)}|V\rangle)
\end{equation}
with
\begin{equation}
\mathrm{tan}(\gamma)=\dfrac{\mathrm{sin(2\theta)}}{t/r-\mathrm{cos}(2\theta)}.
\end{equation}
If $t,r$ are properly chosen such that $t/r=1+\delta$ with $\delta\ll 1$, then $\gamma\approx A\cdot 2\theta$ with $A=1/\delta$ is the factor of amplification. The amplified phase signal can be easily extracted by performing measurement on the basis $\lbrace |R\rangle, |L\rangle\rbrace$ with $|R\rangle=(|H\rangle+i|V\rangle)/\sqrt{2}$ and $|L\rangle=(|H\rangle-i|V\rangle)/\sqrt{2}$, which is completed via polarizing analyser consists of quarter wave plate (QWP), half wave plate (HWP), polarizing beam splitter (PBS) and two electrophotonic detectors (D1 and D2). The output reads
\begin{equation}
\dfrac{I_{1}-I_{2}}{I_{1}+I_{2}}=|\langle R|\phi\rangle|^{2}-|\langle L|\phi\rangle|^{2}=\mathrm{sin}[2(\gamma-\theta)],
\end{equation}
where $I_{1}$ and $I_{2}$ are intensities of light detected by D1 and D2 respectively. Once the amplified phase signal $2(\gamma-\theta)$ is obtained, the gravitational-wave signal can be directly inferred.

The quantum-inspired gravitational-wave detector, which operates based on quantum weak measurements amplification, acts like a hearing aid amplifying phase signal such that gravitational-wave signal can be more easily extracted from the background noise. With the current technologies and facilities, at least two order of magnitude amplification is feasible. In the case of Advanced LIGO, DC readout is applied such that the light intensity from the asymmetric port of Michelson interferometer is
\begin{equation}
I_{A}=I_{in}\cdot[(k\bigtriangleup l)^{2}+2k\bigtriangleup l\cdot\theta+o(\theta^{2})]+I_{d},
\end{equation}
where $\bigtriangleup l$ is the length difference between two arms and $I_{d}$ represents leaking light intensity because of imperfection of interference. Compare readout Eq. (6) to Eq. (5), it is obvious that the quantum inspired gravitational-wave detector is more robust to technical imperfections.

The ultimate sensitivity of ground-base gravitational-wave detectors is mainly limited by the quantum noise. In order to reduce quantum noise in the LIGO, power recycle and Fabry-Perot cavity are used to enhance the input power and arm length. The dual power recycle used in the quantum inspired gravitational-wave detector makes the input power be enhanced more than the case of LIGO. With the method of optical field analysis \cite{analysis}, we obtain gains $G_{p}\approx (1+R)/T=2/T$ and   $G_{arm}\approx 2/\sqrt{\tilde{T}}$, where $T$ and $\tilde{T}$ represent transitivity of the power recycle mirror and input test mirror respectively.
The minimum detectable strain of gravitational-wave, limited by photon shot noise, is thus given by
\begin{equation}
h_{min}\approx \dfrac{1}{(2A-1)G_{p}G_{arm}kL} \sqrt{\dfrac{\hbar\omega}{(I_{1}+I_{2})\tau}}\approx \dfrac{T}{4(2A-1)kL} \sqrt{\dfrac{\tilde{T}\hbar\omega}{(I_{1}+I_{2})\tau}} .
\end{equation}
If parameters above are taken according to the case of LIGO, we obtain $h_{min}\approx 10^{-25}$, which implies improvement of sensitivity with one order of magnitude.

In summary, the quantum-inspired gravitational-wave detector is introduced and shown has potential to improve the strain sensitivity of gravitational-wave detection one order of magnitude further with current technologies and facilities. 

Acknowledgement:
This work was supported by the National Natural Science Foundation of China (No. 11674306 and No. 61590932), the Strategic Priority Research Program (B) of the Chinese Academy of Sciences (No. XDB01030200) and National key R$\&$D program (No. 2016YFA0301300 and No. 2016YFA0301700).

\end{document}